# Designing dislocation-driven polar vortex networks in twisted perovskites

W. Sandholt[1,*], N. Gauquelin[2,*,‡], J. Mangeri[1*], E. Dollekamp[1], G. Panchal[1,2], T. Chennit[2], A. De Backer[2], A. Annys[2], N. Vitaliti[1], A. R. Insinga[1], J. M. Hansen[1], R. Mandal[1], D. R. Rodrigues[3], S. van Aert[2], K. I. Wurster[1], A. Bhowmik[1], I. E. Castelli[1], S. B. Simonsen[1], T. S. Jespersen[1], R. D. James[4], B. Jalan[5], J. Verbeeck[2], J. M. G. Lastra[1], N. Pryds[1,‡]

[1] Department of Energy Conversion and Storage, Technical University of Denmark; Anker Engelunds Vej 1, Kongens Lyngby 2800, Denmark

[2] EMAT and Nanolight Center of Excellence, Department of Physics, University of Antwerpen; Groenenborgerlaan 171, 2020 Antwerpen, Belgium

[3] Department of Electrical Engineering, Politecnico di Bari, Via Edoardo Orabona 4, Bari 70126, Italy

[4] Department of Aerospace Engineering and Mechanics, University of Minnesota-Twin Cities; Minneapolis 55455, Minnesota, United States of America

[5] Department of Chemical Engineering and Materials Science, University of Minnesota-Twin Cities; Minneapolis, 55455, Minnesota, United States of America

[*] These authors contributed equally to this work

[‡] Corresponding authors. Email: nicolas.gauquelin@uantwerpen.be, nipr@dtu.dk

**Twisting two atomic layers produces a geometric moiré pattern, but bonding-induced interfacial reconstruction fundamentally transforms this into an ordered dislocation network - a distinction obscured in weakly-bonded van der Waals systems. Although in-plane topological vortex nanostructures arising from twisting-induced lateral strain modulation have been linked to periodic moiré patterns in freestanding perovskite layers [1, 2, 3] and 2D bilayers [4] , their coupling to the interfacial dislocation network in twisted layers remains unresolved. Here we demonstrate that twisted freestanding SrTiO₃ layers undergo interfacial reconstruction into a network of screw dislocations, accompanied by the emergence of in-plane topological vortices. Unlike in previous reports, these vortices are**

**associated with the periodicity of the dislocation network rather than with geometric moiré patterns. Four-dimensional scanning transmission electron microscopy (4D-STEM) reveals long-range ordered vortex–antivortex arrays with nearly continuous polarisation rotation. A machine learning interatomic potential, trained on first-principles calculations, together with phase-field modelling, confirms that competing strains within the dislocation network stabilize polar vortex–antivortex pairs and drive the emergence of an electronic superlattice with a well-defined periodicity. Our results establish twist-controlled dislocation networks as a new and versatile route to designing local polar and electronic structures in oxide materials.**

Dislocations are topological line defects central to materials science [5, 6]. By relieving elastic energy through local symmetry breaking, they strongly modify mechanical, electronic and structural properties. In two-dimensional materials, ordered defect networks and moiré structures show that such controlled symmetry breaking can generate emergent electronic and optical states [7, 8, 9, 10, 11, 12], establishing dislocations as tunable building blocks for new functionalities.

Although two-dimensional materials provide a powerful platform for moiré engineering, they are mostly limited to compounds with non-covalent interlayer bonding, in which coupling is mediated by weakly hybridized *s* and *p* electrons. This inherently restricts the range of accessible interactions and properties. By contrast, transition metal oxides, governed by strongly correlated *d* electrons, host a much richer electronic landscape absent from most conventional two-dimensional materials. Their strong coupling between charge, spin and orbital degrees of freedom gives rise to phenomena including high-temperature superconductivity [13], colossal magnetoresistance [14], Mott insulating states [15] and multiferroicity [16].

Transition-metal oxides are traditionally grown epitaxially [17, 18], where lattice matching and coherent interfaces restrict viable material combinations. Quasi-atomically thin freestanding oxide membranes now provide an alternative route to synthetic heterostructures through layer-by-layer stacking [3, 19, 20, 21]. By bypassing the constraints of heteroepitaxy, this approach enables interfacial configurations and material pairings that are otherwise thermodynamically or kinetically inaccessible. It also permits control of the relative in-plane orientation between stacked membranes, creating twisted interfaces and in-plane moiré patterns [2, 3, 22].

Stacking two freestanding membranes with a twist angle $\theta$ produces a geometric moiré superlattice with periodicity $\lambda = 2a/\sin(\theta/2)$, where $a$ is the cubic lattice constant [23]. In two-dimensional materials, long-period moirés at small $\theta$ relax into stacked domains separated by dislocation networks [24, 25], and a similar network has been observed in twisted Si [26, 27, 28]. Because coherent interfacial regions are energetically favored, lattice strain drives dislocation formation at small twist angles, whereas larger angles promote preferred coincidence-site-lattice (CSL) configurations [29]. As illustrated in **Fig. 1a**, the interfacial energy at small angles scales with dislocation density, while at larger angles discrete coincidence-site-lattice configurations produce local energy minima. Dislocation formation is therefore expected to be favorable below ~15° [30]. However, direct experimental evidence for such networks in freestanding transition-metal-oxide membranes, and for their effects on structural, physical and electronic properties (**Fig. 1b**), has remained lacking [31].

We show experimentally and theoretically that twisted freestanding $SrTiO_3$ membranes reconstruct into an ordered interfacial network of polar screw dislocations. HAADF and 4D-STEM, supported by *ab initio* image simulations and a DFT-trained machine-learned interatomic potential, reveal a long-range ordered dislocation pattern with polar vortex–antivortex arrays in the interface plane. Each moiré plaquette forms a locally coherent interface, while strain is accommodated by an orthogonal square network of screw dislocations. Phase-field modelling based on atomistic strain gradients shows that polarisation localizes along the dislocation lines, with in-plane nodal chirality and an out-of-plane component. Differential phase-contrast 4D-STEM confirms a spatially periodic internal electric field consistent with this polar texture. DFT further indicates that the reconstructed interface is chiral and hosts mid-gap states absent in bulk $SrTiO_3$, with possible implications for electronic and optoelectronic properties.

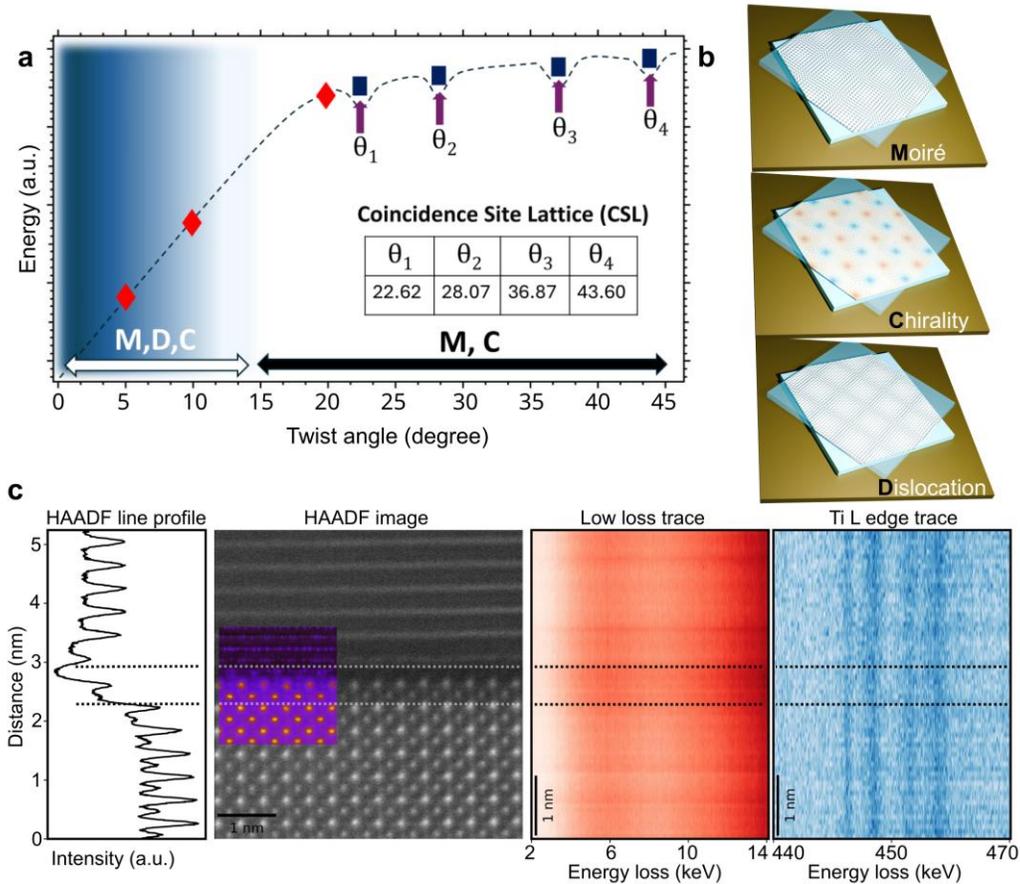

**Fig. 1: Dislocation formation in TMOs and high-quality interfaces. a** Schematic of moiré, chirality, and dislocation regimes as well as the coincident site lattices (CSLs) of the twisted system (blue squares). The red diamond symbols in the figure represent the twist angles explored in this study. **b** Graphical depiction of emergent properties at the interface: moiré pattern (M), chirality (C) and dislocations (D). The regime in which these effects occur is shown in **a**. **c** From left to right: HAADF intensity line profile, cross-section STEM image of twisted $\theta = 20°$, sample showing a clean and sharp STO/STO interface, highlighting the SrO-TiO$_2$ termination, and EELS measurement spectra highlighting the homogeneity and the cleanliness of the interface. The inset in the second panel of **c** is the simulated STEM image based on the structural model described in the text.

## Growth and design of high-quality twisted bicrystals

Freestanding bilayers of STO were created by stacking two STO membranes with twist angles of 5°, 10°, and 20°. **Fig. 1c** (left) shows the HAADF intensity line profile along the [001] direction across a representative 20° twisted STO/STO interface. The interplanar distance $d(001)$ between consecutive Sr atomic columns is approximately $\approx 0.39$nm. Also displayed in **Fig. 1c** is the cross-section HAADF image. These data show that the interface is clean, bonded and free of detectable contamination, with a well-defined [010] zone-axis orientation. The agreement between the cross-sectional STEM image and the corresponding simulation further indicates the absence of unintended defects or inhomogeneities (see inset). Chemical analysis by electron energy-loss spectroscopy (EELS) (two right panels of **Fig. 1c**) confirms the high-quality interface, shown by minimal changes in the Ti L-edge and low-loss EELS across the interfacial plane. Cross-sectional HAADF imaging displayed in **Extended Fig. 1** demonstrates that the interface is well-bonded in the low-angle twist ($\theta = 5°$) sample. See **Methods** and **Supplementary Note 1** for growth, release and stacking along with characterization details (**Supplementary Figs S1-3**).

**Observations of moiré patterns and predicted atomistic reconstruction**

We studied three STO/STO homo-bilayers with twist angles of 5°, 10° and 20°. Dislocations are expected below the critical angle and therefore in the 5° and 10° samples (**Fig. 1a**). In contrast, the 20° sample is above this threshold and does not contain dislocations (see **Fig. 1a** and **Extended Figure 2c**). **Fig. 2a** shows plan-view HAADF-STEM images acquired for the 5° and 20° samples. We also conducted depth-sectioning HAADF-STEM, initially focusing on the top surface of the stack (defocus = 0 nm), followed by an examination of the buried interface (defocus = −12 nm; see **Supplementary Figs. S4-5**). These measurements are additionally supported by abTEM simulations (see **Methods** and **Supplementary Fig. S6**). **Fig. 2b** displays the microscopic moiré structures, which correspond to images simulated using two rigidly rotated (100) STO crystals.

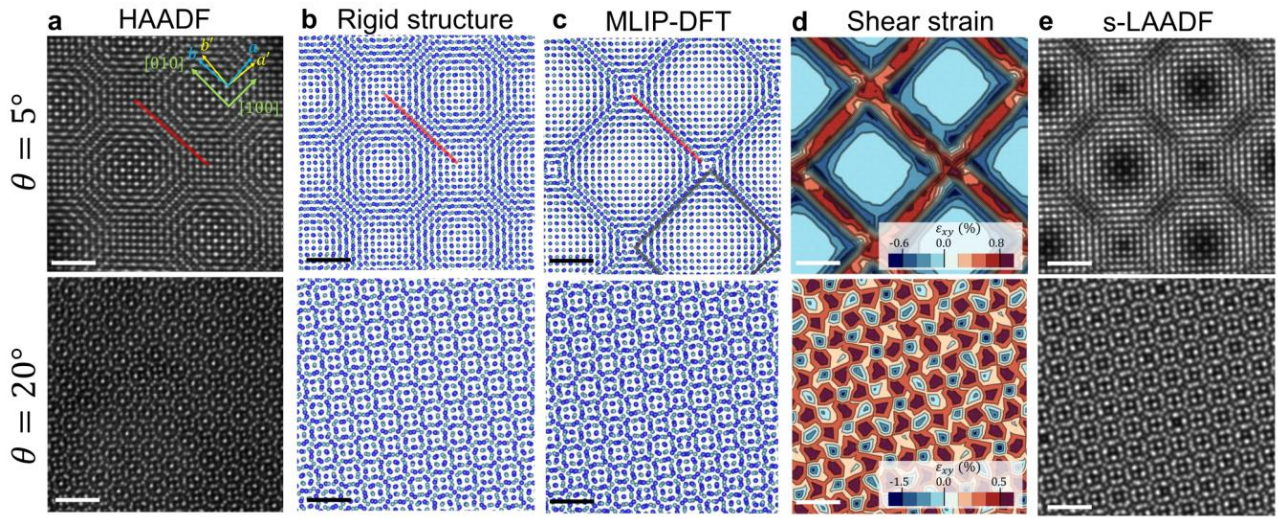

**Fig. 2: Atomistic reconstruction of moiré patterns** (a) Plan-view HAADF-STEM images showing moiré patterns in STO/STO bilayers with 5° and 20° twist angles. (b) Rigid structure rotations of two [001]-oriented STO crystals. Sr and Ti atoms are shown in blue and cyan, respectively. (c) MLIP-DFT predicted relaxed structures showing the dislocation network (black square). (d) Calculated in-plane shear strain component, $\varepsilon_{xy}$ from the relaxed atomic positions at the TiO$_2$-termination in (c). (e) Plan-view image simulated in the low-angle annular dark-field (LAADF)

mode using the atomic positions in (c). All scale bars are 2 nm. The red line in the top panel corresponds to the moiré pattern wavelength $\lambda$ = 4.5 nm. The panel rows correspond to the $\theta$ = 5° and 20°, respectively.

DFT is commonly used to model electronic properties of moiré structures, especially at large twist angles, but its computational cost scales cubically with the number of atoms. To overcome this, we developed an MLIP trained on DFT data (see **Methods** and **Supplementary Note 2, Figs. S7-10**). The experimental atomic reconstructions of the relaxed SrO-TiO$_2$ interface (see **Fig. 1c**) are visualized using the MLIP in **Fig. 2c** for 5° (top) and 20° (bottom) samples. The 5° case shows a clear depiction of two orthogonal screw dislocations oriented parallel to the unstrained cubic reference lattice vectors of the coherent square regions, forming a node at their intersection (see **Extended Figure 2a**). In the 20° case, a crossover is observed to an atomic configuration that is devoid of these strong dislocation features (see estimates of dislocation density in **Extended Figure 2c-e**), albeit still showing a strong similarity to the rigid moiré pattern in **Fig. 2a** and **b**. The reconstructed relaxed interface maximizes Ti coordination, favoring fully coordinated Ti sites (CN=6) by minimizing undercoordinated Ti regions (CN=5) [32] (see **Extended Figure 3**).

Notably, the experimental HAADF-STEM images and the moiré pattern in **Fig. 2a,b** differ from the relaxed MLIP structure predicted for a 5° twist (**Fig. 2c**). This discrepancy mainly reflects the imaging conditions. At a convergence angle of 30 mrad, the electron beam has a depth of focus of about 4 nm, so the signal includes not only the interface but also several layers above and below it, spanning roughly ten layers in total. This through-thickness averaging makes the contrast appear more rigid, consistent with **Fig. 2a,b**. Since moiré patterns can arise even without direct bonding between the two lattices [23, 33], they alone do not constitute reliable evidence of interfacial registry. As a result, well-aligned columns away from the interface can dominate the HAADF intensity maxima, shifting the apparent peak positions and obscuring the true interfacial atomic coordinates. Unless the electron-beam depth of focus is centered exactly on the interface, focused HAADF-STEM images alone can therefore misrepresent the reconstructed structure. Such images are difficult to interpret and can be misleading, particularly at low convergence angles. Consistently, the local interfacial coordination predicted by the MLIP–DFT simulations is far richer than expected from the geometric moiré alone (**Extended Data Fig. 3**).

MLIP–DFT calculations predict a strongly heterogeneous strain landscape that does not simply follow the rigidly twisted moiré geometry, although it shares the same periodicity. The strain tensor of the relaxed structure was estimated from the atomic displacement field (see also **Extended Figure 4**). **Fig. 2d** shows the in-plane shear component, $\varepsilon_{xy}$, at the TiO$_2$-terminated interfacial frame. For both twist angles, sharp gradients in, $\varepsilon_{xy}$, closely follow the square structural motifs of the reconstructed interface (**Fig. 2c**). In the 5° sample, the strain map tracks the square screw-dislocation network, with symmetric shear-strain cores along the dislocation lines. In contrast, the square centres remain coherent and weakly strained. This shows that the strain arises from interfacial reconstruction and dislocation formation. To our knowledge, such a dislocation-driven shear network in oxides cannot be achieved by epitaxial mismatch or externally applied stress.

For comparison with earlier work [1, 34], we also estimated strain directly from the HAADF focal series (**Supplementary Note 3** and **Supplementary Figs. S11-12**). This analysis yields a square vortex-like strain pattern with the periodicity of the moiré lattice, consistent with previous reports [1, 2, 34], but it differs markedly from the calculated interfacial strain profile governed by the dislocation network. This discrepancy highlights the need to resolve the moiré structure at the interface itself; otherwise, the measured strain need not reflect the true interfacial strain state.

Low-angle annular dark-field (LAADF) STEM, which is sensitive to beam channelling, complements HAADF imaging because it enhances strain contrast arising from small atomic displacements [35]. Simulated LAADF images for the 5° and 20° structures, based on the relaxed models in **Fig. 2c**, show dark contrast along the dislocation lines, further supporting the predicted interfacial dislocation network (**Fig. 2e** and **Supplementary Fig. S13**). However, as shown in **Supplementary Fig. S14**, acquiring experimental LAADF images precisely at the interface remains challenging, as with HAADF, due to the limited depth of focus of current convergent-beam instruments.

**Confirming the dislocation networks through cross-section STEM**

To verify the square dislocation network, we examined the twisted bicrystal cross-sections along the orthogonal (100) and (110) planes. These cuts probe different directions through the same interfacial network and therefore provide an internal check of its geometry (**Fig. 3a–c**). The MLIP model predicts projected dislocation spacings of 4.5 nm along (100) and 3.2 nm along (110) (**Supplementary Fig. S14**). Consistent with this, both cross-sectional views show periodic dark-contrast features at the expected dislocation positions: 4.5 nm in **Fig. 3d** and 3.2 nm in **Fig. 3e**.

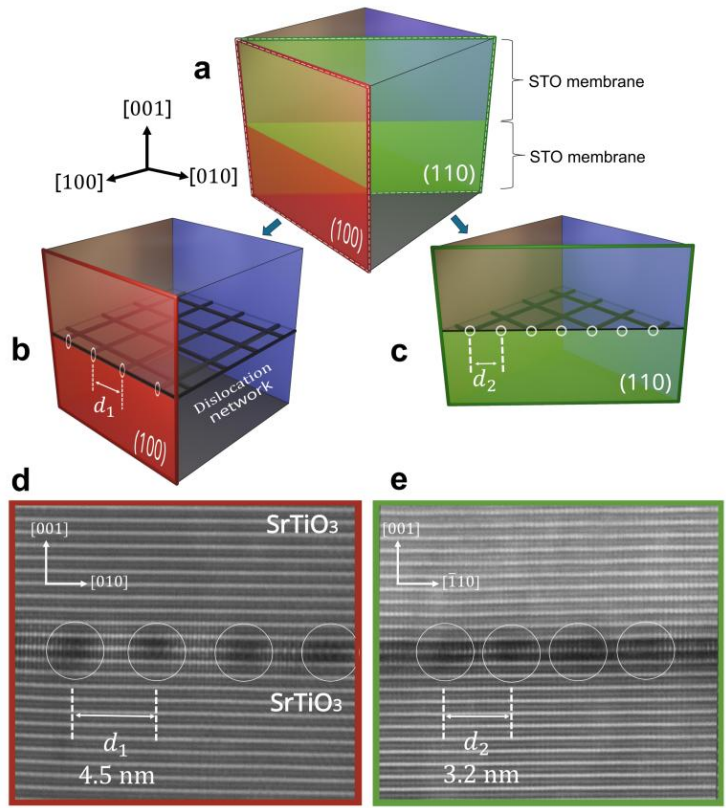

**Fig. 3: Cross-sectional imaging of the dislocation network** (a) Schematic of the two STO membranes oriented in the cubic lattice coordinate system. Illustration of cross-section cuts in the (100) in red, (b) and (110) in green, (c) planes, respectively, highlighting the square moiré plaquette network in black. The (100) and (110) cuts in cross-sectional HAADF imaging are presented in (d) and (e), highlighting the periodicity of dislocations. The distances between image contrast, marked by white circles, are denoted as $d_1$ and $d_2$ corresponding to 4.5 nm and 3.2 nm, respectively.

Pixel-intensity analysis (**Supplementary Fig. S15**) yields spacings of $d_1 = 4.0 \pm 0.3$ nm in **Fig. 3b** and $d_2 = 3.0 \pm 0.3$ nm in **Fig. 3c**, in excellent agreement with the MLIP predictions. The recurring dark contrast is consistent with strain at the dislocation cores and lines, which lowers HAADF intensity through local disorder, lattice tilt and enhanced dechannelling. Such a regular dislocation array is unlikely without a sharp, strongly bonded interface [33]. Consistent with this, the reconstruction is confined to a few atomic planes, and the dark contrast in **Fig. 3d,e** arises from the reconstructed interface rather than imaging artefacts (**Supplementary Fig. S16**). The measured spacings differ by approximately √2, exactly as expected for a *square lattice* viewed along (100) and its diagonal (110), providing independent experimental confirmation of a square interfacial dislocation network.

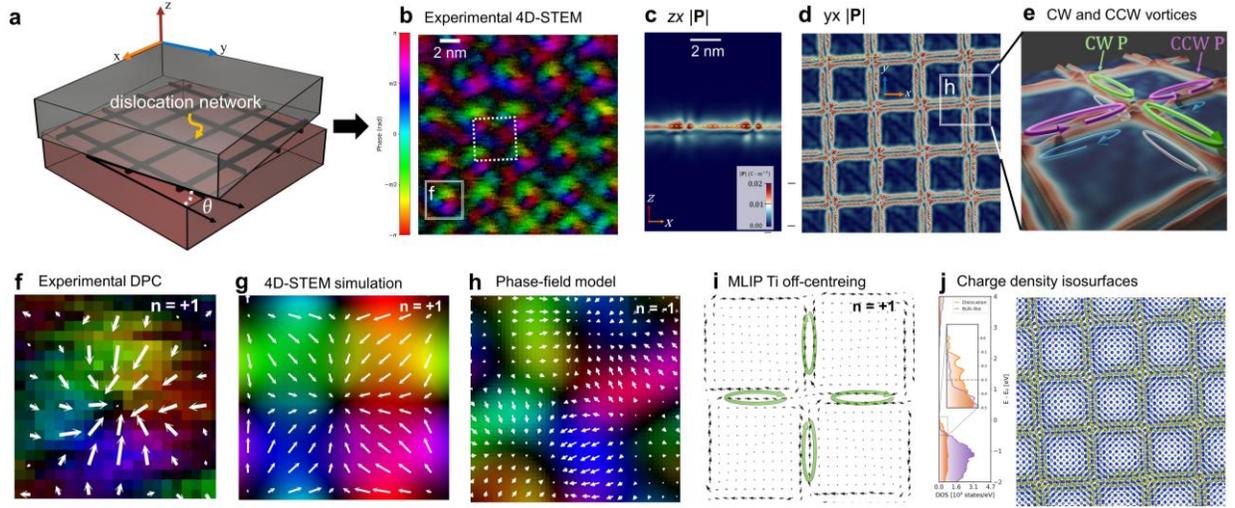

**Fig. 4: Polar textures at screw dislocation cores in twisted SrTiO₃.** (a) Schematic illustration of the twisted oxide layers and their dislocation network. (b) 4D-STEM DPC deflection map of the twisted membrane showing the dislocation location (white dashed square). Color indicates deflection direction; brightness indicates magnitude. The white square corresponds to panel f. (c) $yz$ cross-section of $|\mathbf{P}|$ from the phase field model, showing polar texture confined to within ~2 nm of the interface and tied to the dislocation network shown in the $xy$ plane in (d). (e) 3D rendering of the phase field polarisation with alternating CW and CCW vortices. (f) Experimental DPC at a dislocation core (white box in (b)) showing winding number $n = +1$. (g) 4DSTEM simulation of the dislocation core region using MLIP-relaxed positions, reproducing $n = +1$. (h) Projected electric field orientation map from the phase field model, under the single-column approximation, showing $n = -1$ (antivortex) centered at a dislocation site. (i) Ti atom off-centring informed by the MLIP-relaxed structure, demonstrating only CW polar vortices (green) present at the dislocation cores, yielding $n = +1$. (j) Partial density of states (PDOS, left) and charge density isosurfaces at $E_v - 0.3$ eV (right), highlighting the dislocation network at a twist angle of 6.7°.

**Emergent properties**

We used STEM, 4D-STEM, abTEM and phase-field modelling to map the in-plane electric field at the twisted interface (**Fig. 4a**). In the 5° sample, the 4D-STEM DPC (Differential Phase Contrast) imaging (see **Methods**) shows a periodic deflection pattern with ~4.5 nm spacing (**Fig. 4b**), consistent with the predicted dislocation periodicity (white dashed square). The ordered strain gradients at the interface (~0.03 nm⁻¹) exceed those from membrane bending [36] or wrinkling [37], entering the regime where flexoelectricity induces polarisation [38]. Phase-field modelling (see **Methods**, **Supplementary Note 4** and **Supplementary Figs. S17–19**) confirms a substantial flexoelectric polarisation ($|\mathbf{P}| \approx 0.02$ C·m⁻²) localized along the dislocation lines (**Fig. 4c,d**), decaying within ~2 nm of the interface (**Fig. 4c** and **Extended Data Fig. 5**). The model predicts polarization vortices at each node whose handedness (CW/CCW) alternates laterally, as required by topological charge neutrality across the periodic directions, and also reverses across the interface between the SrO- and TiO$_2$-terminated planes (**Fig. 4e**). The resulting projected electric field, in the single column approximation, carries antivortex winding ($n = -1$) at the dislocation cores (**Fig. 4h**).

Experimentally, however, each core exhibits $n = +1$ winding (**Fig. 4f**), reproduced by abTEM simulation (**Fig. 4g**). The MLIP-relaxed structure reveals the origin: the twisted interface is chiral, hosting only CW Ti off-centering displacements at the cores (**Fig. 4i** and **Supplementary Fig. S20**) rather than the alternating CW/CCW pattern found by the continuum model. This $n = +1$ chirality, imposed by the screw character of the dislocations, breaks the mirror symmetry that would otherwise enforce equal populations of both handednesses, selecting a single winding sense. Capturing this in the continuum description may require an antisymmetric gradient coupling - the electric analogue of the Dzyaloshinskii–Moriya interaction in chiral magnets [39] - absent from the standard flexoelectric functional. The DPC signal contains contributions from both the macroscopic depolarisation field and the atomic-scale potential gradients along distorted columns [40]; because SrTiO$_3$ is nominally paraelectric, the flexoelectric contribution may be insufficient to dominate over the latter, though disentangling the two remains an open experimental challenge. The 20° twist lacks the sharp dislocation network of the 5° case, and the polarisation is correspondingly weaker ($|\mathbf{P}| \sim 0.005$ C·m$^{-2}$). Nevertheless, the phase-field model predicts a rich topological texture of alternating CW/CCW vortices with handedness reversal across the interface (**Supplementary Fig. S19**).

DFT calculations on the relaxed MLIP–DFT structure show that for a 6.7° interface, the occupied states within 0.3 eV of the valence-band maximum ($E_v$), localize near the highly strained screw-dislocation lines (**Fig. 4j**). Partial density-of-states analysis (PDOS, left) further shows that these states (orange) lie *above* the bulk-like valence bands (purple) of less strained regions, and by decomposing the dislocation states into node and edge-line contributions confirms their concentration at the dislocation network (**Extended Figure 6**). Thus, the ordered network may support emergent electronic behaviour in addition to the electromechanical properties shown in Fig. **4a–i**.

**Discussion and outlook**

We present a defect-by-design approach: tuning interlayer commensurability drives self-assembly of targeted defect structures. Twisted freestanding membranes reconstruct into a square, ordered polar dislocation network—its periodicity set by the twist angle rather than a rigid moiré. Instead, they reconstruct into a square, ordered, polar dislocation network whose periodicity is set by the twist angle, providing a directly tunable structural parameter. Unlike in previous reports, the observed vortex–antivortex arrays are linked to the periodicity of the dislocation network, not to a geometric moiré pattern. Interfacial termination (e.g., SrO–SrO vs $TiO_2$–$TiO_2$) is also decisive, as different terminations stabilize distinct local bonding and thus distinct interfacial responses.

The interface reconstruction produces large strain gradients that drive strong flexoelectric polarization, which localizes along dislocation edge lines and extends several nanometres out of plane, imprinting the interface's electromechanical response into adjacent layers. 4D-STEM differential phase-contrast measurements reveal a non-trivial in-plane internal electric-field texture at the dislocation cores, consistent with finite polar order in nominally paraelectric $SrTiO_3$. The charge density at dislocation lines and nodes differs markedly from the bulk, highlighting the reconstructed interface as a platform for emergent electronic phenomena.

These results show reconstructed interfacial defects crucially determine polar chirality and electronic states in twisted membranes. More broadly, they suggest a bottom-up route to functional layered materials and devices, in which nanoscale control of twist angle and termination programs dislocation networks and the coupled strain, polarisation, topology and emergent electronic properties they host.

**Acknowledgements:**

N.P. acknowledges support by the ERC Advanced (NEXUS, grant no. 101054572), a research grant (VIL73726) from Villum Fonden and ERC Synergy Grant METRIQS (no. 101167432). W.S. and J.M. acknowledge resources on the Niflheim supercomputer at the Technical University of Denmark supported by the Novo Nordisk Foundation Data Science Research Infrastructure 2022 Grant: A high-performance computing infrastructure for data-driven research on sustainable energy materials, grant no. NNF22OC0078009. A. A. and J. V. acknowledge the IMPRESS project that has received funding from the Horizon Europe framework program for research and innovation under grant agreement no. 101094299. T. C. and J. V. acknowledge funding from the Flemish Science Fund (FWO) G013122N 'Advancing 4D STEM for atomic scale structure property correlation in 2D materials'. N. G. and J. V. acknowledge funding from the EU's Horizon Europe research and innovation programme under grant agreement no. 101130652 (RIANA). B.J. and R.D.J. acknowledge support from the AFOSR MURI through Grant # FA9550-25-1-0262.

**Author contributions**: NP, NG, JM and JMGL supervised and conceived the project. WS, JMGL, JM, DRR, JMH, ARI, AB, IEC, and ARI performed and analyzed the theoretical calculations. Sample preparation by ED, RM and NV. STEM measurements were performed by NG, SVA, JV and JMH; WS, RDJ, NG, BJ, JM, and NP wrote and revised the article. All authors discussed the results and revised the paper.

**Competing Interests**: The authors do not declare any competing interests.

**Data Availability:** All data supporting the findings of this study will be hosted on a public repository and are made available without restriction.

**Methods:**

**Sample Preparation:** The STO films, each 10 nm thick, were grown on a treated $TiO_2$-terminated STO (001)-oriented substrate with a 10 nm sacrificial layer of $Sr_3Al_2O_6$ (SAO) using high-energy electron diffraction (RHEED) - assisted pulsed laser deposition (PLD) (Coherent KrF UV laser, wavelength 248 nm). The SAO layer was grown at a temperature of 750 °C with a laser fluence of 1.7 J/cm$^2$ and an $O_2$ partial pressure of $9\times10^{-6}$ mbar. The STO film was grown on SAO at 700° C with a laser fluence of 1.3 J/cm$^2$ and $O_2$ partial pressure of $4.6\times10^{-4}$ mbar. Both STO and SAO are grown with a pulse repetition rate of 2 Hz with a chamber base pressure of $1\times10^{-8}$ mbar. To fabricate the membranes, a cellulose acetate butyrate (CAB) support is spin-coated onto the STO/SAO/STO templates. The samples were then immersed in deionized water, which initiated the dissolution of the underlying SAO layer. Once the SAO layer was completely dissolved, the STO membranes, supported by the CAB film, spontaneously detached from the substrate. The membranes were transferred under ambient conditions using a semi-automatic rotation stage, preserving clean interfaces while allowing precise control of the twist angle via a stage manipulator. The stage controller provides < 0.01° step resolution and ±0.1 twist accuracy enabling samples with θ = 5.0°, 10.0°, and 20.0°. The twisted STO stacks are annealed externally in a tube furnace at 700° C for 2 hours in a $H_2/N_2$ (5% $H_2$) atmosphere, with a heating/cooling rate of 2.5° C/min, to ensure interfacial bonding. The membrane stacks are then placed on ultrathin SiN$x$-based TEM grids and analyzed by high-resolution STEM. For cross-sectional imaging, focused ion beam (FIB) milling was used to prepare cuts along the [100] and [110] directions through the bottom membrane, from the part of the TEM grid outside the window where both layers are present. The lamellas were prepared using a dual-beam FIB instrument after the surface of the bilayer film was coated with a 20 nm of carbon in a Leica EM ACE600. See **Supplementary Note 1** for characterisation details.

**Electron Microscopy:** Scanning transmission electron microscopy (STEM) imaging studies were performed at room temperature using the Thermofisher Themis Z 60-300 microscope operated at an acceleration voltage of 300 kV, which is an aberration-corrected microscope equipped with an S-Corr aberration corrector for the probe forming lens and a high-brightness field-emission electron source (X-FEG). Plan view samples and cross-sectional lamellas were studied. For HAADF imaging, a probe size of approximately 0.6 Å with a convergence angle of 30 mrad, and an HAADF collection semi-angle of 64-200 mrad were used. For the LAADF mode, the collection angle was 32-64 mrad using the DF4 detector. The focus was carefully varied in 2-nm increments for focus-dependent measurements, and the pixel size was around 6 pm. All acquired images are drift-corrected using image series and rigid alignment. The cross-sectional images were acquired at 300 kV at a 20 mrad convergence angle and a 60–120 mrad collection angle. Electron energy loss spectroscopy (EELS) was acquired on a Titan 60-300 microscope at 120kV using monochromated STEM-EELS with a convergence angle of 19 mrad, a collection angle of 65 mrad, a pixel size of 0.1 Å, and a dispersion of 25 meV/pixel in dual EELS mode. The Zero Loss Peak (ZLP) was subtracted using a premeasured ZLP in vacuum to give the low-loss EELS signal. The core Loss EELS spectrum of the Ti L edge is treated using background subtraction and binning.

4DSTEM measurements were performed on a Thermo Fisher Scientific Titan 80-300 Double Corrected electron microscope at 300 kV in nano-beam STEM mode with a 1 mrad convergence angle. Scans of 128 by 128 probe positions were performed with a 500 μs dwell time. Diffraction patterns were recorded on a Quantum Detectors MerlinEM Quad direct electron detector with 512 by 512 pixels, mounted on a Gatan Quantum Energy Filter. A slit of 2 eV for zero-loss filtering of the diffraction patterns was employed. From each diffraction pattern, the deflection of the direct beam due to the local electric field is measured. The nanobeam conditions were chosen to average the atomic electric fields within a unit cell and to increase sensitivity to the interface-induced electric field. Data processing was performed using custom-built software that spatially maps the center-of-mass shift of the direct beam.

**Simulated TEM:** The simulated-STEM images were obtained using the abTEM Python package version 1.0.5 [1]. The electron beam energy was set to 300 keV, with a probe convergence semi-angle of 30 mrad. For simulating the interfacial image, a flexible annular detector spanning 30-64 mrad was used. Nyquist sampling was applied, and a focused, non-aberrated probe was used throughout the simulation. For plan-view image simulations, a spherical aberration coefficient of $C_s = 0.159$ µm was applied to the probe. The probe scan sampling was 0.24 Å, and the electrostatic potential was sampled at 0.03 Å/pixel. The simulated potentials were constructed from interface structures obtained from MLIP-DFT relaxed structures, which were combined with rigid $SrTiO_3$ layers to give a total sample thickness of 20 nm. The structure was centered along the beam direction with 4 Å, of vacuum added above and below. The multislice atomic potentials were calculated using Lobato parameterization, a slice thickness of 2 Å, and the infinite projection integral method. The STEM simulations were performed on a single moiré supercell with periodic boundary conditions, allowing the images to be tiled in post-processing to generate a larger field of view. For cross-section images, the interface was cut out of the larger structure and repeated to achieve a total thickness of approximately 50 nm in the beam direction. The [110] orientation was sampled by repeating the interface eight times along the [010]-direction and rotating the cell such that the [010]-axis became the beam direction, while the [100] orientation was generated using a cut along [110]. This was repeated twelve times to yield a total thickness of approximately 54 nm. In the cross-section simulations, periodic boundary conditions were applied only in the lateral directions, not along the beam direction, enabling in-plane tiling of the images. Both the plan-view and cross-section images include an interpolation with a Fourier-space padding of 0.2 pixels/Å, followed by Gaussian blurring with a standard deviation of 0.5 Å, to better match the experimental contrast. Poisson noise was added using an electron dose of $4.2 \times 10^6$ e$^-$/Å, consistent with the experimental imaging conditions. A flexible annular detector spanning 0-200 mrad with a 1 mrad step size was used to generate the LAADF and HAADF detector ranges of (32-120) and (60-200) respectively. For the 4DSTEM simulations, a 1 mrad convergence angle is used with 1 Å Gaussian smoothing. The independent-atom model is used to construct the total potential as a superposition of isolated neutral-atom potentials for both rigid and relaxed structures.

**Machine-learning interatomic potential:** The DFT relaxations, which were used for the training data, were performed at selected twist angles $\theta$ with small and computationally tractable moiré supercells from the method of Refs. [2, 3] (**Supplemental Materials Fig. S7**) using the Vienna Ab initio Simulation (VASP) software [4] with the atomic simulation environment (ASE) [5]. The local density approximation (LDA) was used with the Hubbard + U correction of +5.3 eV applied to the Ti $d$ orbitals [6] within the projector-augmented wave formalism employing a plane-wave cutoff of 500 eV. The forces on each atom in the training set were relaxed to less than 0.03 eV/Å.

An automated batch active learning framework for iterative uncertainty guided data collection and model retraining (CURATOR) [7] is used to generate a comprehensive dataset. An ensemble of three PaiNN models [8] were used, over four active learning iterations, for the uncertainty quantification necessary for the molecular dynamics (MD) sampling in CURATOR. To maximize model accuracy, the MACE equivariant graph neural network architecture [9] is trained on the produced dataset, yielding an accurate MLIP with a mean absolute error of 0.45 meV/atom (see Supplementary Fig. S9). For the results presented, we relaxed the atomic structure while scanning the relative in-plane translation at each $\theta$ to identify the lowest-energy alignment. We refined the final structures by averaging atomic positions from MD simulations at T = 300 K (after discarding the first 10 ps of the 50 ps total equilibration), starting from the most stable translation (**Supplementary Fig. S9**). The in-plane strain tensor components at the interface are estimated from the equilibrated atomic positions using a local reference frame defined by nearest-neighbor averaging, which removes the rigid-body rotational contribution due to the twist. The electronic structure calculations shown in **Fig. 4j** and **Extended Figure 6** are obtained with FHI-aims [10] for the relaxed MLIP-DFT atomic positions, using a linear combination of atomic orbitals in the light basis setting. Scalar relativistic corrections are included with the ZORA atomic expansion and a Hubbard correction of +U = 4.3 eV on the Ti $d$ orbitals (ensuring numerical consistency between the two DFT codes used). The atomic structures displayed in this work are visualized with VESTA [11]. We refer the reader to **Supplementary Note 2** for expanded details on this approach.

**Phase field model:** The 3D phase field modeling employed in this work follows the approach outlined in Ref. [12], which accounts for the flexoelectric coupling in cubic perovskites. The total free energy density includes bulk, gradient, and elastic contributions as well as the coupling to the electric field. The polarisation, elastic displacement, and electrostatic potential are solved self-consistently at each time step through the time-dependent Landau-Ginzburg-Devonshire equation, mechanical equilibrium, and the Poisson equation, respectively. The fourth-order Landau parameterization for STO is taken from Ref. [13] and measurements of the flexoelectric coupling strength are found in Ref. [14]. The strains due to the reconstruction, as estimated by the MLIP-DFT model, are fit to a truncated 2D Fourier series (N = 6) and used as eigenstrain inputs in the phase field simulation of the bilayer stack (total thickness 20 nm). These inhomogeneous strain fields are then reconstructed to ensure mechanical compatibility. Periodicity is enforced on all system variables (elastic displacement vector and polarisation vector components, as well as the electrostatic potential) in the *xy* coordinates. A decomposition method [15] is used to ensure *xy-periodic* strain tensor components. The orientation of the material tensors of the model (e.g. elastic stiffness, electrostrictive, and flexoelectric tensors) is rotated ±θ/2 about the global *z* for each layer in the stack.

The initial condition on each polarization component is assumed to be random and small ($< 10^{-4}$ C/m$^2$). The time-dependent evolution equation is evolved until the change in total energy between adjacent time steps is less than $10^{-6}$ aJ. The finite element calculations are performed within Ferret [16], an open-source module for studying ferroelectric nanostructures, within the Multiphysics Object-Oriented Simulation Environment (MOOSE) [17]. To compare to the 4DSTEM measurements, the self-consistent internal electric field is integrated along *z* in the single-column approximation of an electron traveling along *z*. We refer the reader to the additional details of the continuum modelling approach in **Supplementary Note 4**.

19

19